# Magnetic and transport anomalies in the compounds, RCuAs$_2$ (R= Pr, Nd, Sm, Gd, Tb, Dy, Ho, and Er)

Kausik Sengupta, S. Rayaprol and E.V. Sampathkumaran[*]
Tata Institute of Fundamental Research, Homi Bhabha Road, Mumbai - 400 005, INDIA

Th. Doert and J.P.F. Jemetio
Technische Universität Dresden, Institut für Anorganische Chemie, Mommsenstrasse 13, D-01062 Dresden, Germany

## ABSTRACT

*The results of dc magnetization, heat capacity, electrical resistivity ($\rho$) and magnetoresistance measurements are reported in detail for the compounds, RCuAs$_2$ for R= Pr, Nd, Sm, Gd, Tb, Dy, Ho, and Er, crystallizing in HfCuSi$_2$-type tetragonal structure, with the aim of bringing out anomalies among 'normal' (that is, other than Ce and Yb) rare-earths. The results establish that all these compounds order antiferromagnetically at low temperatures with deviations from de Gennes scaling. Isothermal magnetization (M) data below respective Néel temperatures ($T_N$) reveal the existence of field-induced metamagnetic-like transitions for most of the compounds (except R= Sm and Gd), whereas in Sm and Gd compounds M varies essentially linearly with magnetic field. With respect to the $\rho$ behavior, there appears to be a subtle difference in the temperature dependence beyond 50 K between Sm on the one hand and the rest on the other. In addition, the unexpected $\rho(T)$ minimum reported recently for R = Sm, Gd, Tb and Dy in the paramagnetic state (around 20 to 30 K) is essentially absent for R = Ho and Er. The results overall reveal that the normal rare-earths in this series present an interesting situation in magnetism, warranting a new theoretical approach to describe the transport behavior in the paramagnetic state particularly in the vicinity of $T_N$.*

## 1. INTRODUCTION

The search for new class of ternary rare-earth (R) intermetallic compounds remains unabated, as further investigations, particularly focusing on Ce and Yb compounds, often revealed a variety of novel properties. The interest in the family of the type RTX$_2$, crystallizing in the ZrCuSi$_2$-type layered tetragonal structure (P4/nmm), has started growing in recent years, as even 'normal' rare-earths (that is, apart from Ce or Yb) also have been found to exhibit interesting anomalies revealing richness in the physics of this class of compounds [1-5]. For instance, in RAgSb$_2$ series, (i) LaAgSb$_2$ has been shown to exhibit charge-density-wave (CDW) orderings around 207 K [2], (ii) DyAgSb$_2$ has been reported [1] to show a series of metamagnetic transitions, and (iii) for R= Y, La-Nd, Sm compounds, clear de Haas-van Alphen oscillations in fields as low as 30 kOe and at temperatures as high as 25 K can be measured [1]. In RCuAs$_2$ [Refs. 5-10], the series of interest in the present article, we have reported [5] that there is a minimum in the temperature (T) dependent electrical resistivity ($\rho$) for R = Sm, Gd, Tb and Dy before the long range magnetic order sets in, which is a puzzle. Subsequently, we have subjected this class of compounds across entire series for normal R (R = Pr, Nd, Sm, Gd, Tb, Dy, Ho, Er) to extensive investigations of magnetization (M), $\rho$, magnetoresistance (MR) and heat-capacity (C) in the temperature interval 1.8-300 K. This article presents these results in detail particularly to establish that the transport properties of this class compounds present an interesting scenario.

## 2. CRYSTALLOGRAPHIC FEATURES

The crystal structure of these compounds, shown in Fig. 1, can be viewed as a derivative of the tetragonal CaBe$_2$Ge$_2$-type structure (Fig. 1, left) as elaborated in Ref. 7 with R occupying a site of tetragonal point symmetry. For the ternary rare-earth compounds RT$_2$X$_2$ crystallizing in the CaBe$_2$Ge$_2$ structure, there are layers of atoms stacked in the sequence, R-X-T-X-R-T-X-T along the c-direction (see Fig. 1, left). The structure contains two types of T-atoms, one centering a tetrahedron of X atoms and the other (sandwiched by R and X layers) occupying a top position above a square of X-atoms to give a square pyramid. If one removes all the T ions belonging to the latter category, then one derives the HfCuSi$_2$ type layered tetragonal structure (Fig. 1, middle).

---

[*] Corresponding Author: sampath@tifr.res.in

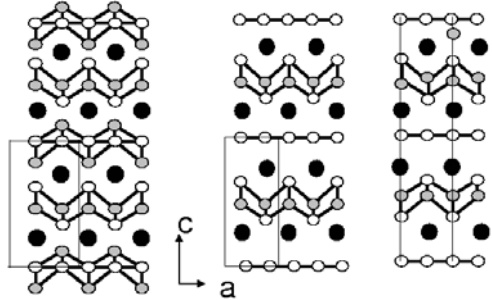

**Fig. 1** *Crystal structures of the CaBe$_2$Ge$_2$ type (left), RCuAs$_2$ (R = Ce - Lu; centre) and LaCuAs$_2$ (right), all projected along [010]. Black circles represent R atoms, small (greyed) circles T (here Be or Cu) atoms and white unfilled circles X (here Ge or As) atoms, respectively. The unit cells are indicated.*

The structures of many binary compounds of the type, RX$_2$, could be viewed as derivatives of the HfCuSi$_2$ structure by removing all T-atoms. It is to be noted that defect structures also are possible with deficiency at the T site or additional ($\delta$) T-atoms occupying the site with square-pyramidal coordination only partially for some partners of R and X. The latter structure is adopted by some RCu$_{1+\delta}$As$_2$ compounds and can be regarded as a 'stuffed variant of HfCuSi$_2$'. Thus, the value of $\delta$ for R = Pr, Nd and Sm in the present series can go up to 0.09, 0.06 and 0.05 respectively (however with a value of 1 for heavier R) in the RCuAs$_2$ series. Finally, it is to be remarked that, among the ternary rare-earth compounds of the type RCu$_{1+\delta}$As$_2$, all but La crystallize in the tetragonal space group P4/nmm with lattice constants of about a = 4 Å and c = 10 Å (number (Z) of formula units per unit cell being 2), while LaCu$_{1+\delta}$As$_2$ crystallizes (Fig. 1, right) in a two-fold supercell space group I4/mmm with doubling of the c-axis with Z = 4 which is due to a different stacking sequence of the La, Cu and As layers (compare middle and right figures in Fig. 1).

3. EXPERIMENTAL DETAILS

The polycrystalline samples RCuAs$_2$ (R= Pr, Nd, Sm, Gd, Tb, Dy, Ho, Er, Y and Lu) were obtained by solid-state method starting from the stoichiometric amounts of the elements in evacuated and sealed silica ampoules. The ampoules were kept in a chamber furnace at 770 K for two days and then at 1170 K for ten days. After furnace cooling to room temperature, the samples were characterized by powder X-ray diffraction (Cu K$_\alpha$) and the lattice constants obtained are given in table 1. The samples are found to be greyish black in colour. For the purpose of obtaining a rod for $\rho$ and C measurements, the powder samples were pelletized and sintered again at 1170 K. The magnetic measurements were performed employing two commercial magnetometers (1.8 - 300 K) - vibrating sample magnetometer (Oxford Instruments) and superconducting quantum interference device (Quantum Design). The $\rho$ data were obtained (1.8-300 K) by a conventional four-probe method and the data was also collected as a function of magnetic-field (H) at selected temperatures. The C measurements (1.8 - 40 K) were carried out but a semi-adiabatic heat-pulse method.

4. RESULTS AND DISCUSSION
A. MAGNETIZATION

The T-dependence of zero-field-cooled (ZFC) magnetic susceptibility ($\chi$) taken in an H of 5 kOe is plotted in Fig. 2 for all samples in the form of inverse $\chi$ versus T.

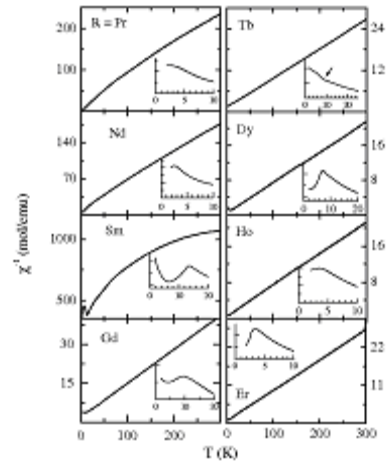

**Fig. 2** *(a) Inverse dc magnetic susceptibility ($\chi$) as a function of temperature (T) for zero-field-cooled specimens of RCuAs$_2$ compounds, measured in the presence of an external magnetic field (H) of 5 kOe. In the insets, the shapes of the $\chi$(T) curves at low temperature are shown to highlight the features due to magnetic ordering in each case. An arrow for Tb case is shown to indicate a change in slope at $T_N$.*

The insets show the data in an expanded form at low temperatures to highlight the features due to magnetic ordering. A common feature for all these compounds, except the Sm case, is that the $\chi$ exhibits Curie-Weiss behavior at high temperatures, say, above 100 K, and at lower temperatures a marginal deviation arises from crystal-field effects. The values of the effective magnetic moment ($\mu_{eff}$) and the paramagnetic Curie temperatures ($\theta_p$) obtained from the linear region are given in table 1.

**Table 1**: *The lattice constants, a and c, the paramagnetic Curie temperature ($\theta_p$), Néel temperature ($T_N$), de Gennes scaled $T_N$ values ($T_N^d$), effective magnetic moment ($\mu_{eff}$, 100- 300 K), and the temperature ($T_{min}$) at which the $\rho$ minimum appears in the paramagnetic state, for $RCuAs_2$ class of compounds.*

| R  | a (Å) | c (Å)   | $\theta_p$ (K) | $T_N$ (K) | $T_N^d$ (K) | $\mu_{eff}$ ($\mu_B$) | $T_{min}$  |
|----|-------|---------|-----------|-----------|-------------|-----------------------|------------|
| Pr | 4.008 | 10.0496 | -51       | 4.1       | 0.46        | 3.5                   | Absent     |
| Nd | 3.979 | 10.0162 | -30       | 2.5       | 1.0         | 3.86                  | weak, 11 K |
| Sm | 3.944 | 9.944   | --        | 13        | 2.55        | --                    | 35 K       |
| Gd | 3.917 | 9.9391  | -15       | 9         | 9           | 7.9                   | 35 K       |
| Tb | 3.901 | 9.9012  | -14       | 8         | 6           | 9.7                   | 35 K       |
| Dy | 3.884 | 9.8472  | -3        | 8         | 4.05        | 10.6                  | 20 K       |
| Ho | 3.868 | 9.8042  | -9        | 4.4       | 2.97        | 10.8                  | Absent     |
| Er | 3.865 | 9.8022  | -6        | 4.0       | 1.46        | 9.5                   | weak, 30 K |

The values of $\mu_{eff}$ are found to correspond to that of respective trivalent R ions. The negative sign of $\theta_p$ indicates antiferromagnetic nature of the exchange interaction. In the case of SmCuAs$_2$, it is well known that the non-linearity in the plot of inverse $\chi$ versus T even at high temperatures arises from the fact that the first excited state of the Hund's rule multiplet (J = 7/2) is very close to the ground state (J = 5/2). For this reason, these parameters are not derived for this sample (and hence not shown in table 1). At low temperatures, there is a distinct feature due to magnetic ordering, revealed either by a peak (for Nd, Sm, Gd, Dy, Ho, and Er) in the $\chi(T)$-plot or a sudden change in slope in $1/\chi(T)$ (for Pr and Tb). These features due to the onset of magnetic ordering thus appear around 4.1, 2.5, 13.0, 9, 8, 8, 4.4 and 4 K for Pr, Nd, Sm, Gd, Tb, Dy, Ho and Er respectively. There is a weak upturn at further temperatures in most cases, which we attribute to the presence of traces of paramagnetic impurities, as the isothermal M curves look somewhat similar at temperatures just above and below these upturns. We would like to mention that YCuAs$_2$ is found to be diamagnetic ($\chi$ (300K) = -0.0002 emu/mol) and LuCuAs$_2$ is essentially a Pauli paramagnet ($\chi$ (300K) = 0.0017 emu/mol). Hence Cu is non-magnetic.

In order to understand the nature of the magnetic ordering, we have obtained isothermal M at selected temperatures in the vicinity of magnetic transition temperatures up to 120 kOe. A common feature to all these compounds is that M is found to be non-hysteretic even at 1.8 K, without getting saturated in the low field range. This finding indicates that these compounds are neither ferromagnets nor spin-glasses. A further careful look at the plots (Fig. 3) in fact confirms that all these compounds should be classified as antiferromagnets as evidenced below.

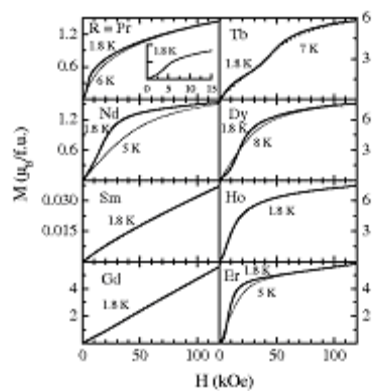

*Fig. 3* *Isothermal magnetization (M) as a function of magnetic field (H) for RCuAs$_2$ compounds at selected temperatures (with the thicker continuous line representing 1.8 K data). In the case of R = Pr, the curve below 20 kOe is shown in an expanded form in the inset to highlight a sudden change in the slope.*

Another common feature for all the compounds except the case of R = Sm and Gd is that, say at 1.8 K, following an initial linear variation of M with H, there is a curvature of the plot of M(H) as though there is a metamagnetic transition, followed by a tendency to saturate at higher fields. The field at which this transition takes place depends on R. Thus, this process takes place around 4, 8, 40, 10, 2, and 3 kOe for Pr, Nd, Tb, Dy, Ho, and Er respectively. This curvature is in the upward direction for all the cases except for Tb compound, in which there is a tendency to saturate in the range 20 to 40 kOe; beyond this range, of course, the upward curvature is noticed; the feature appearing for the intermediate field range gets diminished as the T is increased, say, to 7 K for Tb compound.

In the case of Sm and Gd, these features are absent; instead, M appears to vary nearly linearly with H till the highest field measured; it is not clear whether the absence of this feature only for these two is due to anisotropic nature of the materials. Nevertheless, all these features can be explained only if one assumes that the zero-field state is antiferromagnetic below respective magnetic ordering temperatures. Incidentally, as in the case of $SmAgSb_2$ (Ref. 1), the value of M is remarkably small even at high fields for $SmCuAs_2$, while the corresponding value for the Gd compound is far less than the free ion value, which is attributable to antiferromagnetism. It is worth noting that, unlike in $DyAgSb_2$ (Ref. 1), multiple steps in the plots of M(H) in $DyCuAs_2$ are absent, which reveal that such effects are not specific for Dy for this structure but the ligands play a major role to decide such magnetic features.

We have also obtained low-field (100 Oe) $\chi$ data for ZFC and field-cooled (FC) states of the specimens (Fig. 4).

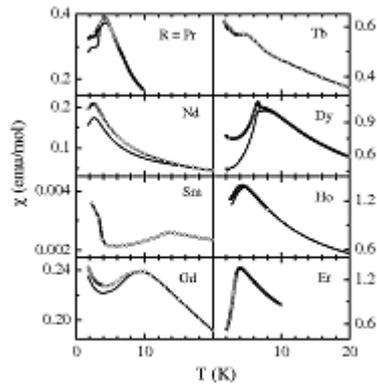

*Fig. 4 Magnetic susceptibility measured in the presence of 100 Oe for the zero-field-cooled (continuous lines) and field-cooled (points) conditions of the specimens, $RCuAs_2$. In the case of Er, both the curves overlap in the entire temperature range.*

As in the case of high field $\chi$ data shown in Fig. 2, there are well-defined features very close to the Néel temperatures ($T_N$) (that is, a peak in all cases, except for $TbCuAs_2$). In the Tb case, one sees only a sudden change of slope around 9 K followed by a peak at a lower temperature. In the case of Dy sample, one sees multiple peaks around $T_N$ in this low-field $\chi$ data (which merge into one in the high-field data, see Fig. 2), thereby revealing interesting field dependence. In some cases (Ho and Er), there is a distinct upward shift of the peak temperature with respect to that in the 5kOe-data by about 0.5 K (Fig. 2), as though the magnetic field suppresses magnetic ordering temperature, consistent with antiferromagnetism of these compounds. The point of emphasis is that the ZFC-FC curves overlap for Sm and Er compounds in the T-range of investigation, while the bifurcation of these curves at $T_N$ is rather weak for the rest of the compounds, which is attributed to domain effects in long range magnetically ordered systems, reported quite frequently in the literature. These results imply that all these compounds are not spin-glasses as evidenced by the absence of hysteresis in isothermal M behavior as well. It is interesting to note that these curves tend to bifurcate, though weakly, at a temperature higher than $T_N$ for Pr (around 6 K) and Nd (around 16 K) compounds. We attribute this to the domain effects from possible short-range order.

Various parameters derived from all these data are listed in Table 1. It may be noted that the values of $T_N$ are in general larger than that expected on the basis of de Gennes scaling. Several factors including anisotropy due to crystal-field effects, and sensitivity of Fermi surface to small changes in lattice constants as discussed for isostructural $RAgSb_2$ compounds [1], in addition to 4f hybridization effects for light rare-earths [11], may be responsible for this deviation. As this is not the main emphasis of this article, we will not dwell on this any further.

**B. HEAT CAPACITY**

The results of C measurements are shown in Fig. 5 to obtain further support for magnetic ordering at low temperatures. There are features attributable to magnetic ordering in the vicinity of the temperature at which $\chi(T)$ shows corresponding anomalies. It is rather difficult to precisely estimate $T_N$ from the C data, as the C(T) curve above the peak is in general broad due to various factors like possible modulations in the magnetic structure, short range order etc [12]. Considering these, one can confidently state that the $T_N$ determined from the C data agrees within 1 K with those obtained from $\chi$ data.

In the case of Pr, the feature in the raw data is not very transparent and the magnetic ordering manifests itself as a change in the slope of C(T) plot, which is attributable to the low entropy associated with the lowest-lying level responsible for magnetic ordering. However, we could not estimate the magnetic entropy ($S_m$) at $T_N$ satisfactorily for this compound due to narrow temperature window of the data below $T_N$. We could however get an inference about the behavior of $S_m$ for other compounds for which $T_N$ is somewhat higher. For this purpose, we obtained [12] the 4f contribution ($C_{4f}$) employing the values of C of YCuAs$_2$ for the lattice contribution.

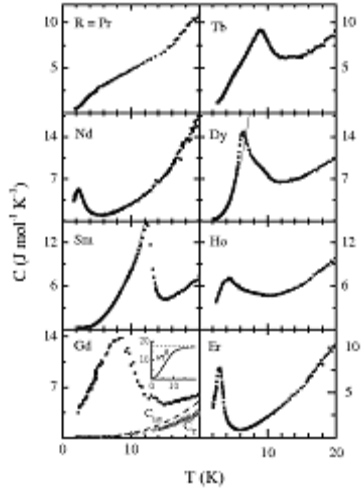

***Fig. 5*** *Heat capacity (C) as a function of temperature (T) below 20 K for RCuAs$_2$ compounds. For Sm and Dy compounds, the continuous line represents a fit to $T^3$ dependence of C. For the Gd case, the lattice contribution ($C_{latt}$) to C derived on the basis of the knowledge of C of Y compound ($C_Y$) is also shown. The magnetic entropy, $S_m$ (in units of J/mol K), derived from $C_m$ data is also plotted in the inset and the horizontal dashed line represents R ln 8.*

To estimate $S_m$ we extrapolated the C data from 2 K to zero Kelvin temperature assuming C is proportional to $T^3$ in the range 0 – 5 K, a justifiable assumption for antiferromagnets, *vide infra*. There could be some ambiguities in the absolute values of $S_m$ at higher temperatures (say, well beyond 10 K), as one is not sure whether the Y compound serves as a good reference for lattice contribution at such temperatures. However, in low temperature range, say below 10 K, at which the phonon contribution is negligible, one can trust the values of $S_m$ for qualitative conclusions. The values of $S_m$ derived at $T_N$ are close to half of that expected for fully degeneracy of the R ions in most cases (Sm, Tb and Dy), while for Gd, at $T_N$, we obtain about 80% of the full value. We illustrate $S_m(T)$ behavior for Gd case in figure 5 and it is clear that the full value (R ln 8) is recovered at a temperature close to $2T_N$ only. While in most cases the reduced value of $S_m$ can be attributed to crystal field effects, an additional factor has to be invoked, considering that the crystal field effects have to be ignored for Gd. This behavior is somewhat similar to that reported for many other Gd alloys, possible implications of which have been discussed at length earlier [12, 13]. It is important to note that C varies as $T^3$ (see the continuous line in Fig. 5) for SmCuAs$_2$ and DyCuAs$_2$ below the peak, which is consistent with the antiferromagnetic ordering. It is difficult to verify this functional dependence for other compounds, due to the fact that either $T_N$ is low (close to the temperature limit of our measurements) or the C(T) plot appears to be complex due to various other factors like crystal-field and Zeeman effects [12]. We would like to add that the values of linear coefficient of C for YCuAs$_2$ and LuCuAs$_2$ are found to be very small (close to 1mJ/mol K$^2$) and the corresponding values of Debye temperature are about 265 and 215 K respectively as inferred from the C data in the range 7 - 15 K.

## C. ELECTRICAL RESISTIVITY

We now focus on the ρ(T) behavior (Fig. 6).

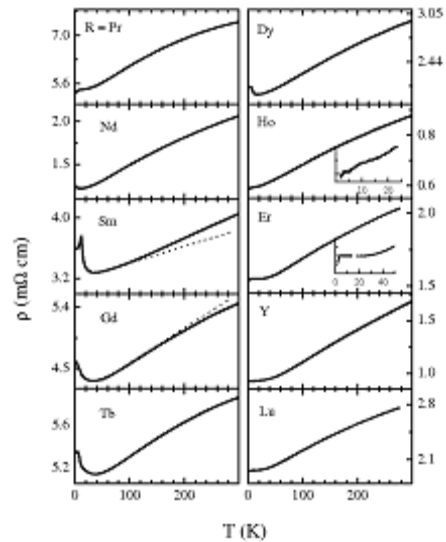

***Fig. 6*** *The electrical resistivity as a function of temperature (1.8 - 300 K) for RCuAs$_2$ (R= Pr, Nd, Sm, Gd, Tb, Dy, Ho, Er, Y and Lu). The data in an expanded form for Ho and Er are plotted in the insets to show that the minimum in the plot is absent for these compounds. In the case of Sm and Gd, the dashed lines are obtained by linear extrapolation of the data in the range 50 to 100 K to highlight the direction of change of slope of ρ(T) curves at higher temperatures.*

Since the samples are porous, the absolute values are overestimated. It is however clear that the values fall in the milliohms range and the temperature coefficient of ρ is positive down to at least 50 K, which imply that these materials behave like metals. There is a positive curvature in the ρ(T) plots in all cases below about 150 K (except for R = Sm) as though there is a tendency for saturation at T >> 300 K; the fact that this is true for Gd case as well implies that this positive curvature is controlled by the phonon contribution rather than the crystal-field effects. At higher temperatures, the plots are essentially linear, which implies that the phonon contribution dominates electrical resistivity in that range. In the case of Sm, ρ increases with T (beyond, say, 100 K) faster than the linearly extrapolated values from lower temperatures. Thus, there appears to be a subtle difference in the transport behavior at high temperatures among these compounds, (see the dashed lines in Fig. 6 for Sm and Gd cases extrapolated from the data in the range 50 - 100 K). The different behavior of the Sm compound can not be due to close spacings of Hund's rule multiplet states (J = 5/2 and 7/2), as one does not see a similar difference in the isostructural $RAgSb_2$ series [1]. The reason beyond this anomaly is however not clear to us at present. As stressed earlier [5], the puzzling finding is that there is a well-defined minimum for R = Sm, Gd, Tb and Dy in ρ(T) at a temperature ($T_{min}$) far above $T_N$ (see table 1), the upturn being in the range of 4 to 14% depending on R, which is very difficult to understand within commonly known concepts. While, we will return for further discussion of this feature in the next paragraph, it is important to note that the minimum is weak for $ErCuAs_2$, in the sense that the magnitude of the upturn below $T_{min}$ (30 K) is about 0.5% only, if one expands the inset further in Fig. 6. In the case of $HoCuAs_2$, the minimum is completely absent. As stated before [5], this feature is absent for R = Pr and Nd. Thus, the transport behavior viewed together within this series is quite fascinating in the entire temperature range.

We will now focus our discussion on $ρ_{min}$ briefly, and it has been discussed at length in our earlier articles [5, 13 -15]. For this purpose, we have also shown the data in various ways for the Sm case in Ref. 5. In order to illustrate our findings and to support our point of view, we show similar plots here for another case, viz., $GdCuAs_2$, (see Fig. 7) in the T region of interest (10 - 30 K) above $T_N$. It is clear that none of the plots is linear. While the formation of any type of gap is excluded from the non-linear plot of ln(ρ) versus 1/T, non-linearity of ln(ρ) versus $T^{-1/4}$, $T^{-1/2}$ and $T^{1/2}$ rules out variable range hopping, Coulomb gap and presently known weak-localization mechanisms [16, 17] respectively.

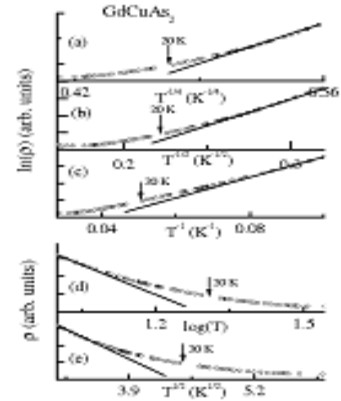

**Fig. 7** *The electrical resistivity in the temperature range 10 - 30 K for $GdCuAs_2$ plotted in various ways. A line is drawn through the points in some range to highlight that all the points do not lie on this line.*

Since the 4f level is deeply localized, one cannot attribute it to the Kondo effect and consistent with this there is no logarithmic dependence of ρ below $T_{min}$. Considering that the non-magnetic compounds, $YCuAs_2$ and $LuCuAs_2$, behave normally in the sense that there is no minimum in the ρ(T) plot (see Fig. 6), $ρ_{min}$ in some other cases within the same series can not arise from band structure effects, but must be magnetic in origin. The ρ anomaly cannot arise from short-range magnetic order alone, as this kind of order also results in a loss of spin-order contribution resulting in a decrease of ρ rather than an increase with decreasing T. However, it is not clear whether this short-range order triggers some kind of weak localization (indirect effect of short range order?), which requires to be addressed theoretically. In short, the origin of the transport anomaly intrinsically lies in magnetism.

**D. MAGNETORESISTANCE**

Since all these compounds are antiferromagnets, one can also make a naive proposal that there is a short range antiferromagnetic order above $T_N$ persisting till $T_{min}$ and that the negative temperature coefficient of ρ in the range $T_N < T < T_{min}$ arises from antiferromagnetic pseudo-gap formation [18,19] as a magnetic precursor effect. This aspect has normally not been considered above $T_N$ in the literature for any system. If this is true, one should see similar gap effects below $T_N$ as well. While the ρ exhibits an increase below $T_N$ for R = Gd and Tb indicative of antiferromagnetic gap formation for these cases, it decreases for the rest of the cases discussed here. However, the magnitude of net increase below $T_N$ for Gd and Tb cases, when the pseudo-gap is expected to have fully formed, is lower than that in the above T range in the paramagnetic state. This does not speak in favor of such pseudo-gap arguments to explain $\rho_{min}$. To get a convincing picture, the investigation of the influence of H on $\rho$ is important. One should see negative MR (defined as MR = $[\rho(H)-\rho(0)]/\rho(0)$) as a function of H with a pronounced variation at low fields in the magnetically ordered state, in the event that there is a pseudo-gap. With this in mind, we carried out MR measurements at 4.2 K for those cases for which $T_N$ is above 4.2 K (R = Sm, Gd, Tb, Dy), in addition to one more temperature (15 K) above $T_N$. The data (with the experimental error in MR being of the order of 0.1%) are shown in Fig. 8.

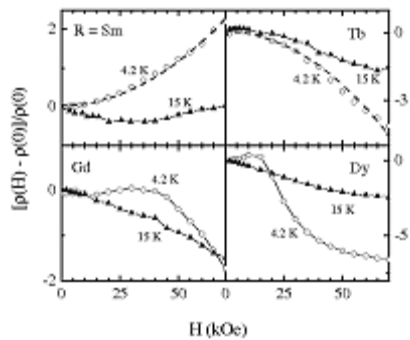

**Fig. 8** The magnetoresistance, $[\rho(H)-\rho(0)]/\rho(0)$, as a function of magnetic field at 4.2 and 15 K for $RCuAs_2$ (R=Sm, Gd, Tb, and Dy). The continuous lines through the data points are guides to the eyes, while the dashed line in some cases represent a fit to a quadratic dependence of MR on H.

It is distinctly clear from Fig. 8 that, in the case of **SmCuAs$_2$**, the values of MR are positive at 4.2 K varying quadratically with H, typical of antiferromagnetic metals without pseudo-gap; at 15 K, the values of MR are very small (comparable to the experimental error) with negligible field dependence, which establishes that there is no pseudo-gap formation above $T_N$ as well. While this example itself is enough to prove the point, it is worthwhile to look at the MR behavior for other cases as well. For the **Gd compound**, MR at 4.2 K is nearly zero till about 40 kOe, and there is no noticeable MR with a negative sign for small fields. Beyond 40 kOe, of course, negative MR appears gradually with H, attributable to the continuous evolution of the ferromagnetic alignment by the applied field. At 15 K, MR varies essentially quadratically typical of paramagnets, without any sharp decrease for initial applications of H. The MR behavior of **TbCuAs$_2$** is somewhat similar to that of Gd compound. In the case of **DyCuAs$_2$**, at 4.2 K, we see a distinct positive sign of MR below 15 kOe, which implies antiferromagnetism without pseudogap at low fields; there is a sign reversal of MR around 20 kOe, which is consistent with field-induced metamagnetic transition inferred from the isothermal M data. At 15 K, $\rho$ decreases rather sluggishly with increasing H as in other members discussed above. Thus the MR behavior of all these compounds cannot be consistently understood if one attributes the $\rho$-minimum to the formation of an antiferromagnetic pseudo-gap at smaller length scales above $T_N$. It may also be remarked that the sluggish variation of MR with H at 15 K rules out possible role of grain boundary effects as a possible cause of $\rho_{min}$.

5. SUMMARY

To conclude, we have reported the results of first detailed measurements of magnetization, specific heat, electrical resistivity and magnetoresistance on a series of compounds, $RCuAs_2$ [20]. The results establish that the moment-containing R ions undergo antiferromagnetic ordering at low temperatures. There is a subtle difference in the high temperature (> 50 K) $\rho(T)$ behavior of Sm system on the one hand and the rest on the other and such findings could be of importance to current theoretical interests to understand high temperature $\rho$ behavior in metals [21]. We have also elaborated on the observation [5] that there is an unexpected prominent minimum in the $\rho(T)$ above $T_N$, qualitatively similar to that observed in Ce-based Kondo lattices, for many of these (R = Sm, Gd, Tb, Dy). The present work, apart from establishing that this cannot be explained within commonly known ideas including possible antiferromagnetic pseudo-gap effects persisting above $T_N$, brings out that such a minimum does not extend for other normal heavy R (e.g., HoCuAs$_2$). This finding makes the situation more puzzling than envisaged earlier [5] and it is not clear whether a small change in the lattice constants brought out by lanthanide contraction crucially controls the (hitherto unknown) mechanism responsible for $\rho(T)$ minimum. Finally, it is worth noting that similar $\rho_{min}(T)$ has been reported recently among the heavy rare-earth members of the series, RAgGe, by Moroson et al [22]. Thus, there are now at least three different classes of materials, $R_2TMX_3$ (TM = Cu, Pd, Pt; X= Si, Ge; Refs. 13, 14, and 15), $RCuAs_2$ and RAgGe, in which such a $\rho(T)$ minimum is distinctly seen for 'normal' heavy rare-earths, apart from weak features observed in other systems as well (Refs. 13, 14, 19). In fact Europium metal also has been known to exhibit a minimum in ρ(T) (around 120 K) above magnetic ordering temperature [23] and in our opinion this transport anomaly in Eu metal also requires a fresh look. As reiterated by Moroson et al, this situation, apparently more widespread as evidenced above, certainly asks more theoretical input in magnetism, particularly to explain the circumstances under which ρ(T) minimum appears for some normal rare-earths, but not for others, within the same series as revealed by the results in this article.

ACKNOWLEDGMENTS

The authors would like to thank Kartik K Iyer for his help while performing the experiments.

[20] We have carried out preliminary studies on $CeCuAs_2$ also. Ce is essentially trivalent and it appears that this compound is a non-magnetic heavy fermion down to 0.5 K with a negative temperature coefficient of $\rho$ in the range 0.5-300 K. The results will be published elsewhere after carrying out further studies.